\documentstyle[epsf,12pt]{article} \begin{document}
\title{Mass dependence of the HBT radii observed in $e^+e^-$ annihilation}

\author{A.Bialas and K.Zalewski\\ M.Smoluchowski Institute of Physics
\\Jagellonian
University, Cracow\thanks{Address: Reymonta 4, 30-059 Krakow, Poland;
e-mail:bialas@thp1.if.uj.edu.pl, Zalewski@chall.ifj.edu.pl}\\
Institute of Nuclear Physics, Cracow}
\maketitle

\begin{abstract}

It is shown that the recently established strong mass-dependence of the
radii of the hadron sources, as observed in HBT analyses of the $e^+e^-$
annihilation, can be explained by assuming a generalized inside-outside
cascade, i.e. that (i) the {\it four}-momenta and the space-time
position {\it four}-vectors of the produced particles are approximately
proportional to each other and (ii) the "freeze-out" times are
distributed along the hyperbola $t^2-z^2= \tau_0^2$.

\end{abstract}

It has been found recently that the parameters describing the B-E
interference in $e^+e^-$ annihilation depend strongly on the masses of the
particles used in the analysis \cite{al,lep}. One finds $r_{\pi}$ between
$0.7$ and 1 fm;$\;
r_K$ between $0.5$ and $0.7$ fm;$\;
 r_{\Lambda}$ between $0.1$ and $0.2$ fm.

 In the
present note we suggest that this dependence can be understood if
the produced particles   satisfy approximately
the (generalized) Bjorken-Gottfried conditions \cite{go,bj}:

(i) The 4-momentum $q_{\mu}$ and the 4-vector $x_{\mu}$ describing the
space-time position of the production ("freeze-out") point of a particle
are proportional
 \begin{equation}
 q_{\mu} =\lambda  x_{\mu}. \label{1}
 \end{equation}
The proportionality factor $\lambda$ is a scalar with respect to boost
in the longitudinal direction;

(ii) Particles are produced at a fixed proper time $\tau_0$ after the
collision, i.e.
 \begin{equation}
t^2-z^2=\tau_0^2 \label{1a}
 \end{equation}
where $t,z$ are time and longitudinal position of the production point.

From (\ref{1}) and (\ref{1a}) we derive
\begin{equation}
\lambda = \frac{M_{\perp}}{\tau_0}           \label{1b}
\end{equation}
where $M_{\perp}^2= E^2 -q_{\parallel}^2 $.
 Thus finally we have
 \begin{equation} q_{\mu} =
\frac{M_{\perp}}{\tau_0} x_{\mu}. \label{2}
 \end{equation}

This picture is, of course, purely classical and can only be treated as
a heuristic guide-line when applied to actual production processes.
 A more
adequate formulation of these conditions  can be achieved
using the Wigner representation $W(P,x)$
of the (single-particle) density matrix
which, as is well known (see e.g. \cite{cz}), corresponds -
as close as possible without contradicting quantum mechanics -
to the space-time and momentum distribution of the produced particles.
To implement the conditions (i), (ii) above, we postulate $W(P,x)$ in
the form
\begin{eqnarray}
 W(P,x) \sim  \delta( t^2-z^2 -\tau_0^2)
 \exp\left[-\frac{x_{\perp}^2}{2R_{\perp}^2}
 -\frac{P_{\perp}^2}{2\Delta_{\perp}^2}\right] \nonumber \\
\exp\left[-\frac{\left(P_+-\frac{M_{\perp}}{\tau_0}x_+\right)^2
+\left(P_--\frac{M_{\perp}}{\tau_0}x_-\right)^2}
{2\delta_{\parallel}^2}-
\frac{\left(P_{\perp}-\frac{M_{\perp}}{\tau_0}x_{\perp}\right)^2}
{2\delta_{\perp}^2}\right]
\label{3}
\end{eqnarray}
where
\begin{equation}
x_{\pm} = t\pm z;\;\;\;\; P_{\pm}= P_0\pm P_z.    \label{3a}
\end{equation}
so that
\begin{equation}
M_{\perp}^2=P_+P_- ;\;\;\;\;\; \tau_0^2=x_+x_-. \label{3b}
\end{equation}
 The first
exponential represents
a standard cylindrically symmetric "longitudinal"
distribution in  momentum and in
configuration space\footnote{To simplify the argument,
 we ignore the longitudinal momentum and $z$ dependence of the
single particle spectrum. This seems a reasonable approximation
at high energy.}.
 The new point is the second exponential which
introduces correlation between the momentum and the point of emission  of
the particle, as required by the generalized
 Bjorken-Gottfried condition (\ref{2}).
Such correlations are known to influence strongly the HBT effect on
particle spectra \cite{bow}.
It is thus this factor\footnote{ Admittedly, the form
(\ref{3}) is rather schematic. In particular, gausssians are taken for
simplicity and can be replaced if necessary.  We also did not include
fluctuations of $\tau_0$. These simplifications are not essential for
our argument, however. } which, we think, is responsible for the mass
dependence of the observed HBT radii.

To derive HBT correlations we need to calculate from (\ref{3}) the
density matrix in momentum space (see e.g.\cite{sh,bk,he,bz}).
 This can be done using the relation
between $W(P,x)$ and $\rho(q,q')$ which reads
\begin{equation}
\rho(q=P+\frac{Q}2,q'=P-\frac{Q}2)= \int d^4x e^{iQx} W(P,x).
  \label{2a}
\end{equation}
From (\ref{2a}) we see that now we have to take
\begin{equation}
 P=\frac{q+q'}2;\;\;\;\; M_{\perp}^2 =P_+P_-;\;\;\;\; Q=q-q' \label{8}
 \end{equation}
with the 4-momenta $q$ and $q'$ on the mass-shell.

To continue, it is convenient, as usual, to introduce the rapidities
\begin{equation}
Y=\frac12 \log\frac{P_+}{P_-};\;\;\;\; \eta =\frac12 \log\frac{x_+}{x_-}
\label{4}
\end{equation}

 The longitudinal integral
\begin{eqnarray}
 I_{\parallel}\equiv
\int d\eta
\exp\left(-\frac{M_{\perp}^2}{2\delta_{\parallel}^2}
\left[\left(e^Y-e^{\eta}\right)^2
+\left(e^{-Y}-e^{-\eta}\right)^2\right]\right) \nonumber \\
\exp\left(i\tau_0\left[m_{\perp}\cosh({y}-\eta)
-m_{\perp}'\cosh({y}'-\eta) \right]\right) \label{9}
 \end{eqnarray}
where $(m_{\perp},{y})$ and $(m_{\perp}',{y}')$ are transverse
masses and rapidities corresponding to momenta $q$ and $q'$, can be
approximated by
 \begin{eqnarray} I_{\parallel}\approx
\int d\eta\exp\left(-\frac{M_{\perp}^2}{\delta_{\parallel}^2}
(Y-\eta)^2\right)
\nonumber \\
\exp\left(i\tau_0\left[ m_{\perp}\left(1+\frac
{(\eta-{y})^2}2\right) -m_{\perp}'\left(1+\frac
{(\eta-{y}')^2}2\right)\right]\right) \label{10}
 \end{eqnarray}
Ignoring normalization and phase
factors, inessential for our argument, we
thus obtain
\begin{eqnarray} I_{\parallel} \sim
 \exp \left(
-i \tau_0 \frac{M_{\perp}^2}{2B\delta_{\parallel}^2}
(m_{\perp}(Y-{y})^2-m_{\perp}'(Y-{y}')^2)\right) \nonumber \\
\exp\left(-\frac{\tau_0^2}{4B} m_{\perp}m_{\perp}'({y}-{y}')^2 \right)
\label{11}
\end{eqnarray}
where
\begin{equation}
B=
\frac{M_{\perp}^2}{\delta_{\parallel}^2}
-i\frac{\tau_0}2(m_{\perp}-m_{\perp}') \label{12}
\end{equation}
 The transverse
integral can be evaluated exactly.
 Ignoring again the normalization and phase factors
we have
 \begin{eqnarray}
I_{\perp}\equiv
\exp\left(-\frac{\vec{P}^2}{2\Delta_{\perp}^2}\right)
\int d^2x \exp\left( -\frac{\vec{x}^2}
{2R_{\perp}^2} - \frac {\left(
\vec{P}-\frac{M_{\perp}}{\tau_0}\vec{x}\right)^2}
{2\delta_{\perp}^2} - i\vec{Q}\vec{x}\right) \nonumber \\ \sim
\exp\left[-\frac{(\vec{q}+\vec{q}')^2}{8\Delta_{eff}^2}
-\frac{(\vec{q}-\vec{q}')^2R_{eff}^2}{2} \right]
 \label{13}
\end{eqnarray}
 where all vectors are two-dimensional (transverse) and
\begin{equation}
\frac1{\Delta_{eff}^2}=
\frac{\tau_0^2}{M_{\perp}^2R_{\perp}^2+\tau_0^2\delta_{\perp}^2}
+\frac1{\Delta_{\perp}^2};\;\;\;\;
 R_{eff}^2=
\frac{R_{\perp}^2\tau_0^2\delta_{\perp}^2}
{M_{\perp}^2R_{\perp}^2+\tau_0^2\delta_{\perp}^2} \label{14}
\end{equation}

From (\ref{13}) we find the  single particle
transverse momentum distribution:
\begin{equation}
\frac{d\sigma}{d^2q_{\perp}}\sim
I_{\perp}(\vec{q}=\vec{q}'\equiv\vec{q}_{\perp})=
\exp\left(-\frac{q_{\perp}^2 \tau_0^2}
{2(m^2+q_{\perp}^2)R_{\perp}^2+2\tau_0^2\delta_{\perp}^2}
-\frac{q_{\perp}^2}{2\Delta_{\perp}^2}\right).
\label{14aa}
\end{equation}
One  sees that the average transverse momentum is largely determined by the
value of $\Delta_{\perp}$ which thus cannot be too large if one wants to
insure average transverse momentum smaller than, say, $500$ MeV.

Let us also note at this point that consistency
 with uncertainty principle implies the inequality \cite{bz}
\begin{equation}
R_{eff}\Delta_{eff} \geq  \frac12    \label{14a}
\end{equation}
As seen from (\ref{14}), at large transverse mass $M_{\perp}$, this
inequality can only be satisfied if $\delta_{\perp}$ is significantly
larger than $\Delta_{\perp}$ (and thus than the average transverse momentum).

To proceed, we shall assume
that all correlations between particles which
are not caused by Bose-Einstein interference can be
 neglected. Using the
formulation of \cite{bz} we thus write the two-particle density
matrix as a product
\begin{equation}
\rho(q_1,q_2;q_1',q_2')=\rho(q_1,q_1')\rho(q_2,q_2')  \label{15}
\end{equation}
It then follows from the general theory of HBT effect
 (see, e.g.\cite{he})
 that the observed two-particle distribution is given by
\begin{equation}
\Omega(q_1,q_2)= \rho(q_1,q_1)\rho(q_2,q_2)+
\rho(q_1,q_2)\rho(q_2,q_1)\equiv\Omega(q_1)\Omega(q_2)(1 \pm C(q_1,q_2))
 \label{16}
\end{equation}
where
\begin{eqnarray}
C(q_1,q_2)= C_{\parallel}(q_1,q_2)C_{\perp}(q_1,q_2) \nonumber \\
= \frac{\mid I_{\parallel}(q_1,q_2)\mid^2}
{I_{\parallel}(q_1,q_1)I_{\parallel}(q_2,q_2)}
\frac{\mid I_{\perp}(q_1,q_2)\mid^2}
{I_{\perp}(q_1,q_1)I_{\perp}(q_2,q_2)}  \label{17}
\end{eqnarray}
 describes the HBT correlations.

Using (\ref{13}) we  find
\begin{equation}
C_{\perp} = e^{ - (\vec{q}_1-\vec{q}_2)^2 R_{HBT}^2} =
e^{ - Q_{\perp}^2 R_{\perp HBT}^2}
\label{19}
\end{equation}
where
\begin{equation}
R^2_{\perp HBT}= R^2_{eff} -\frac1{4\Delta_{eff}^2}= \frac{\tau_0^2}
{M_{\perp}^2 R_{\perp}^2 + \tau_0^2 \delta_{\perp}^2}
 (R_{\perp}^2\delta_{\perp}^2-\frac14)- \frac1{4\Delta_{\perp}^2}  \label{20}
\end{equation}
Since
\begin{equation}
M_{\perp}^2 = \left( \frac {m_{1\perp} +m_{2\perp}}2\right)^2 +
m_{1\perp}m_{2\perp} \sinh^2\left(\frac{y_1-y_2}2\right)   \label{18}
\end{equation}
we conclude that indeed $R^2_{\perp HBT}$ falls with increasing (transverse)
mass of the particle.

For $C_{\parallel}$ we have
\begin{equation}
C_{\parallel}(q_1,q_2) = \exp\left(-R_{\parallel HBT}^2(m_{1\perp}y_1-m_{2\perp}y_2
+(m_{1\perp}-m_{2\perp})Y)^2\right)\label{21}
\end{equation}
where
\begin{equation}
R_{\parallel HBT}^2 = \frac{\tau_0^2 M_{\perp}^2}{2|B^2| \delta_{\parallel}^2}
\label{22}
\end{equation}
From  (\ref{12}), one sees that also $R_{\parallel HBT}^2$ falls
 with increasing $M_{\perp}^2$.

If, as is customary (see e.g. \cite{al}), one works in the frame where
$Y=0$, (\ref{21}) can be written as
\begin{equation}
C_{\parallel}(q_1,q_2) \approx e^{-R_{\parallel HBT}^2(q_{1z}-q_{2z})^2}
=e^{-R_{\parallel HBT}^2Q^2_{\parallel}}     \label{23}
\end{equation}

This completes the qualitative discussion of the mass effect in our
approach. It remains to be seen if the values of the HBT radii given by
(\ref{20}) and (\ref{22}) can be adjusted to be close
 to the ones obtained from the LEP data \cite{al,lep}.

\begin{figure}
\epsfxsize 6cm
\begin{center}
~\epsfbox{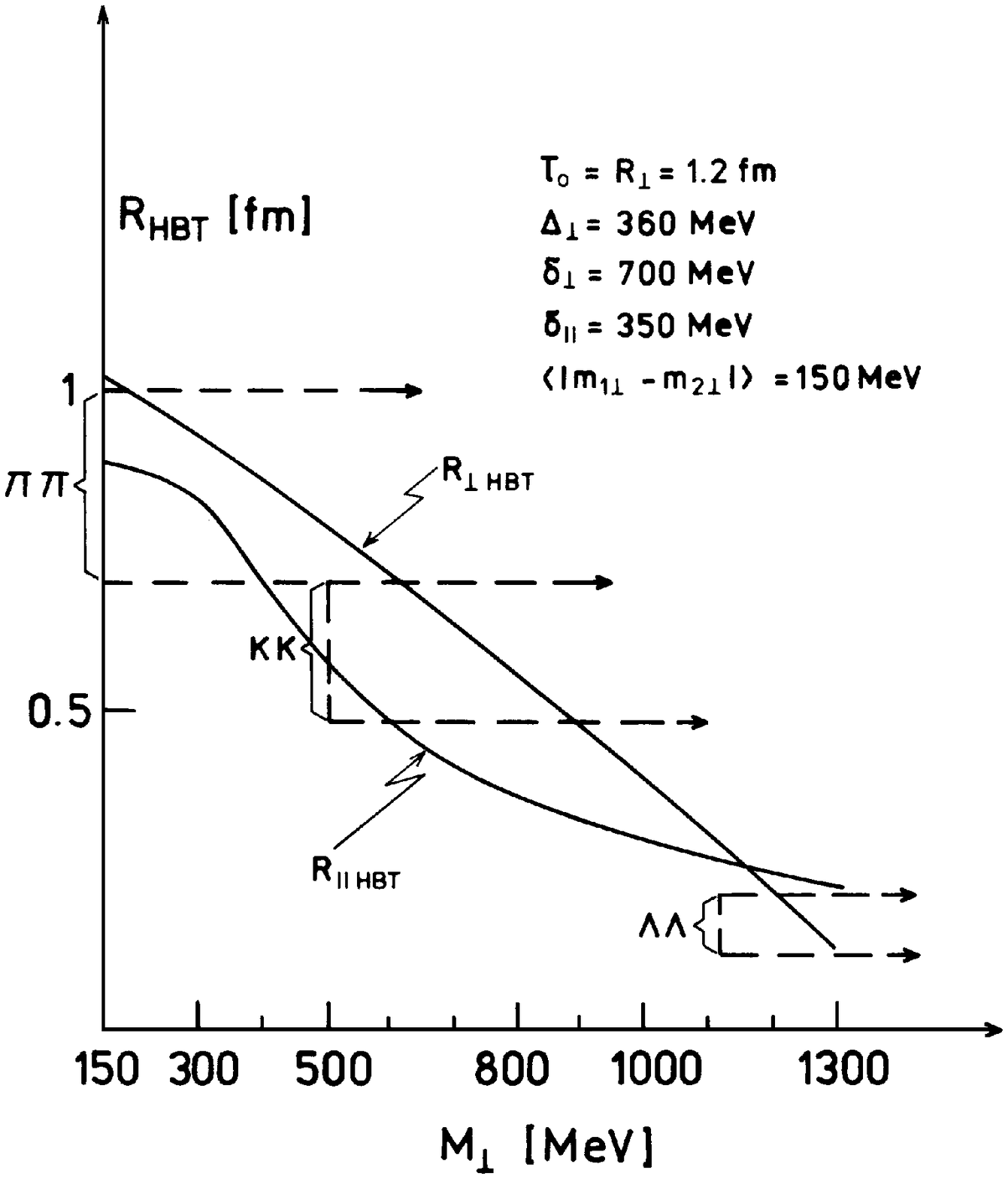}
\end{center}
Figure 1. $R_{\perp HBT}$ and $R_{\parallel HBT}$ plottet versus
$M_{\perp}$. The parameters are shown in the figure. The data
from $\pi\pi$, $KK$ and $\Lambda\Lambda$ correlations are also
indicated.
\end{figure}

In Fig. 1 $R_{\parallel HBT}$ and $R_{\perp HBT}$ are plotted versus
$M_{\perp}$, the transverse mass of the two-particle system. The values
of other parameters were taken as follows : $\Delta_{\perp}$ = 360 MeV,
$\tau_0= R_{\perp} = 1.2$ fm, $\delta_{\perp}= 700$ MeV,
$\delta_{\parallel} = 350$ MeV, $|m_{1\perp}-m_{2\perp}| = 150$ MeV. One
sees a rather strong mass dependence of both longitudinal and transverse
radii. We did not try to fit the obtained values to the data as this
would require working directly with  data themselves and thus goes
beyond the scope of the present investigation. It is nevertheless
recomforting to observe that the HBT radii, obtained with
"reasonable" values of the model parameters, are not far from the ones
found in LEP experiments.

We thus conclude that the existing data on HBT radii are consistent
with the hypothesis
 that -in $e^+e^-$ annihilation at high energy- 4-momentum of a
produced particle is approximately proportional to its space time
position 4-vector at the freeze-out time\footnote{Recently an alternative
interpretation has been proposed in \cite{ALX}.}.

This proportionality is of course well-known for the {\it longitudinal }
components \cite{go,bj}, and is exhibited
explicitly in numerous models \cite{models}.
At this point our approach is similar to the one proposed for a
longitudinally expanding fireball \cite{cs,he}, although the mass
dependence following from our Eq. (\ref{3}) seems somewhat stronger. On
the other hand, a rather novel feature following from our analysis is
that the original Gottfried-Bjorken proportionality relation should be
extended to include also the {\it transverse} components of the
4-vectors, as explicitely expressed in (\ref{2}).

Several comments are in order.

(i) It should be emphasized that our argument is only semi-quantitative
and can be improved in many details when applied to real data. In
particular, the gaussians in the Wigner function (\ref{3}) can be
replaced by more realistic functions for numerical analysis. Also, the
Fourier transform (\ref{9}) can be calculated numerically without
approximations shown in (\ref{10}), which were introduced simply to
obtain an analytic result. Finally, including a distribution of
$\tau_0$ is probably needed to obtain a good description of data. We
feel, however, that all this necessary fine tuning does not invalidate
our main conclusion, summarized in Eq. (\ref{2}).

(ii) As we already mentioned, the results shown in Fig. 1 do not
represent a fit to experimental data which we think would be premature
at the present stage. Therefore, the values of the parameters used to produce
this figure are by no means final. Some of them seem rather stable,
however. In particlular, $\Delta_{\perp}$ is closely related to the average
transverse momentum and thus cannot be arbitrarily changed. Also a
rather large value of $\delta_{\perp}$ seems necessary to satisfy the
consistency condition (\ref{14a}).
This means that the correlation between the tranverse
momentum and transverse position of a particle at freeze-out is fairly
weak. It is remarkable that such a weak correlation is sufficient to
create a strong variation of $R_{\perp HBT}$ with the transverse mass of
the investigated two-particle system.

(iii) From the point of view of data analysis, our argument emphasises
the importance of the investigation of the HBT correlations as function
of the transverse mass of the pion pair.

(iv) Relation (\ref{2}), when applied to transverse directions, implies
the existence of an important "collective transverse flow" in the system
of particles produced in $e^+e^-$ annihilation\footnote{It roughly
corresponds to Bjorken's proposal of "expanding shell" \cite{bj} (c.f. also
\cite{pra}.}. It
would be interesting to search for other evidence of such a "flow" in
the data.

(v) A natural modification of the relation (\ref{1a}) is to consider
freeze-out times given by the fully Lorentz-invariant formula
\begin{equation}
t^2-z^2-y^2-x^2 =\tau_0^2   \label{x}
\end{equation}
which leads to qualitatively similar results as those discussed in the
present paper. It is not clear if the present data can distinguish
between (\ref{1a}) and (\ref{x}) but investigation of this question
is  certainly  a challenging issue for future work.

(vi) The recent data of L3 coll. \cite{l3} show a strong dependence of
the transverse $\pi\pi$ HBT radius (and a somewhat weaker dependence of
the longitudinal radius) on the average transverse mass of the two pions
$m_{\perp}= \frac12 (m_{1\perp}+m_{2\perp})$. This seems not
inconsistent with our results, although more work is needed to establish
a closer connection between $M_{\perp}$ and the average trasverse mass
$m_{\perp}$ which is used to parametrize the data. Thus before more
detailed calculations (including a realistic single particle
distribution) are performed, it is not clear to what extent the results
shown in Fig.1 are related to the observations of \cite{l3}.

\vspace{0.3cm}
{\bf Acknowledgements}
\vspace{0.3cm}

We would like to thank G.Alexander for calling our attention to this
problem and for informing us about his recent results. The help of
H.Palka in interpretation  of data is also highly appreciated.
This investigation was supported in part by the KBN Grant No 2 P03B 086
14.


\begin{thebibliography} {99}
\bibitem{al}
G. Alexander and I.Cohen, Tel-Aviv preprint, hep-ph/9811338, to be
published in Proc. of "Hadron Structure '98" (Stara Lesna, Sept. 1998).
\bibitem{lep}
 DELPHI coll., P.Abreu et al.,
 Phys.Letters B286 (1992) 201; Phys. Letters 379 (1996) 330; Proc. EPS
Conf, Jerusalem (1997) and H.Palka, private comm.

OPAL coll.,P.D.Acton et al. Phys.Letters B298 (1993) 456;
 G.Alexander et al., Phys. Letters B384 (1996) 377; Z.Phys. C73 (1996) 389;

 ALEPH coll.,  D.Decamp et al., Z.Phys. C54
(1992) 75; D.Baskulic et al., Z.Phys. C64 (1994) 361;
\bibitem{go}
K. Gottfried, Phys.Rev.Letters 32 (1974) 957; Acta Physica Polonica B3
(1972) 769; F.E.Low and K.Gottfried, Phys.Rev D17 (1978) 2487.
\bibitem{bj}
J.D.Bjorken, Proc. SLAC Summer Inst. on Particle Physica, SLAC-167
(1973) Vol.I, 1;
Phys.Rev D7 (1973) 282; Phys.Rev. D27 (1983) 140; Proc. of
the XXIV Int. Symp. on Multiparticle Dynamics, Vietri (1994), ed. by A. Giovannini et al,
World Scientific 1995 ,p.579.
\bibitem{cz}
P.Carruthers and F.Zachariasen, Rev. Mod. Phys. 55 (1983) 245.
\bibitem{bow}
 M.G.Bowler, Z.Phys. C29 (1985) 617.
Y.M.Sinyukov, in Hot Hadronic Matter: Theory and Experiment, eds.
J.Lettesier et al., Plenum , NY, 1995 p. 309.
W.A. Zajc, in Particle Production in Highly Excited Matter, eds.
H.H.Gutbrod and J.Rafelski, NATO ASI Series B 303 (1993) 435.
\bibitem{sh}
E.Shuryak, Phys. Lett. 44B (1973) 387; S.Pratt, Phys.Rev. Lett. 53
(1984) 1219.
\bibitem{bk}
A.Bialas and A.Krzywicki, Phys.Letters B354 (1995) 134.
\bibitem{he}
S.Chapman, P.Scotto and U.Heinz, Heavy Ion Physics 1 (1995) 1 and
references quoted there; K.Geiger et al., hep-ph/9811270.
\bibitem{bz}
A.Bialas and K.Zalewski, Eur.Phys.J. C6 (1999) 349; Phys. Letters B436
(1998) 153.
\bibitem{models}
X.Artru and  G.Menessier Nucl.Phys. B70 (1974) 93; B.Andersson et
al.,Phys.Reports 97 (1983) 31; J.Kogut and L.Susskind Phys. Rep. 8C
(1973) 85.
\bibitem{ALX}
G. Alexander, I. Cohen and E. Levin, Tel Aviv preprint TAUP-2549-99 and
hep-ph/9901341.
\bibitem{cs}
A.Makhlin and Y.Sinyukov, Z.Phys. C39 (1988) 69;
T.Csorgo, Phys. Lett. B347 (1995) 354.
\bibitem{pra}
S.Pratt, Phys.Rev.Letters 53 (1984)13;
 T.Csorgo and B.Lorstad, Phys. Rev. C54 (1996) 1390.
\bibitem{l3}
L3 Coll., L3 98-2268 Proc. ICHEP98 Vancouver (1998) Ref. 506.


\end{thebibliography}
\end{document}